\documentclass[11pt]{article}
\pdfoutput=1
\usepackage{hyperref}
\usepackage{graphicx}
\usepackage{epsfig}
\usepackage{epsf}
\usepackage{epstopdf}

\begin{document}
\pagestyle{empty}
\def\eqa{\!\!&=&\!\!}
\def\ccr{\nonumber\\}

\def\la{\langle}
\def\ra{\rangle}

\def\del{\Delta}
\def\ddel{{}^\bullet\! \Delta}
\def\deld{\Delta^{\hskip -.5mm \bullet}}
\def\ddeld{{}^{\bullet}\! \Delta^{\hskip -.5mm \bullet}}
\def\dddel{{}^{\bullet \bullet} \! \Delta}

\def\rld{\rlap{\,/}D}
\def\rldd{\rlap{\,/}\nabla}
%
\def\half{{1\over 2}}
\def\third{{1\over3}}
\def\fourth{{1\over4}}
\def\fifth{{1\over5}}
\def\sixth{{1\over6}}
\def\seventh{{1\over7}}
\def\eigth{{1\over8}}
\def\ninth{{1\over9}}
\def\tenth{{1\over10}}
\def\bN{\mathop{\bf N}}
\def\R{{\rm I\!R}}
\def\Eins{{\mathchoice {\rm 1\mskip-4mu l} {\rm 1\mskip-4mu l}
{\rm 1\mskip-4.5mu l} {\rm 1\mskip-5mu l}}}
\def\Z{{\mathchoice {\hbox{$\sf\textstyle Z\kern-0.4em Z$}}
{\hbox{$\sf\textstyle Z\kern-0.4em Z$}}
{\hbox{$\sf\scriptstyle Z\kern-0.3em Z$}}
{\hbox{$\sf\scriptscriptstyle Z\kern-0.2em Z$}}}}
\def\abs#1{\left| #1\right|}
\def\com#1#2{
        \left[#1, #2\right]}
\def\square{\kern1pt\vbox{\hrule height 1.2pt\hbox{\vrule width 1.2pt
   \hskip 3pt\vbox{\vskip 6pt}\hskip 3pt\vrule width 0.6pt}
   \hrule height 0.6pt}\kern1pt}
      \def\boxop{{\raise-.25ex\hbox{\square}}}
\def\contract{\makebox[1.2em][c]{
        \mbox{\rule{.6em}{.01truein}\rule{.01truein}{.6em}}}}
\def\ltap{\ \raisebox{-.4ex}{\rlap{$\sim$}} \raisebox{.4ex}{$<$}\ }
\def\gtap{\ \raisebox{-.4ex}{\rlap{$\sim$}} \raisebox{.4ex}{$>$}\ }
\def\mn{{\mu\nu}}
\def\rs{{\rho\sigma}}
\newcommand{\Det}{{\rm Det}}
\def\Tr{{\rm Tr}\,}
\def\tr{{\rm tr}\,}
\def\sumij{\sum_{i<j}}
\def\e{\,{\rm e}}
\def\pa{\partial}
\def\dA{\partial^2}
\def\ddx{{d\over dx}}
\def\ddt{{d\over dt}}
\def\der#1#2{{d #1\over d#2}}
\def\lie{\hbox{\it \$}} 
\def\partder#1#2{{\partial #1\over\partial #2}}
\def\secder#1#2#3{{\partial^2 #1\over\partial #2 \partial #3}}
%
\newcommand{\be}{\begin{equation}}
\newcommand{\ee}{\end{equation}\noindent}
\newcommand{\bear}{\begin{eqnarray}}
\newcommand{\ear}{\end{eqnarray}\noindent}
\newcommand{\benn}{\begin{enumerate}}
\newcommand{\enn}{\end{enumerate}}
\newcommand{\veject}{\vfill\eject}
\newcommand{\ven}{\vfill\eject\noindent}
%
\def\eq#1{{eq. (\ref{#1})}}
\def\eqs#1#2{{eqs. (\ref{#1}) -- (\ref{#2})}}
%
\def\totint{\int_{-\infty}^{\infty}}
\def\posint{\int_0^{\infty}}
\def\negint{\int_{-\infty}^0}
\def\pint{{\dps\int}{dp_i\over {(2\pi)}^d}}
%
\newcommand{\GeV}{\mbox{GeV}}
\def\FFdual{F\cdot\tilde F}
\def\bra#1{\langle #1 |}
\def\ket#1{| #1 \rangle}
\def\braket#1#2{\langle {#1} \mid {#2} \rangle}
\def\vev#1{\langle #1 \rangle}
\def\rightvac{\mid 0\rangle}
\def\leftvac{\langle 0\mid}
\def\ihbar{{i\over\hbar}}
\def\ge{\hbox{$\gamma_1$}}
\def\gz{\hbox{$\gamma_2$}}
\def\gd{\hbox{$\gamma_3$}}
\def\go{\hbox{$\gamma_1$}}
\def\gt{\hbox{\$\gamma_2$}}
\def\gth{\hbox{$\gamma_3$}} 
\def\gf{\hbox{$\gamma_5\;$}}
\def\slash#1{#1\!\!\!\raise.15ex\hbox {/}}
\newcommand{\slD}{\,\raise.15ex\hbox{$/$}\kern-.27em\hbox{$\!\!\!D$}}
\newcommand{\slpartial}{\raise.15ex\hbox{$/$}\kern-.57em\hbox{$\partial$}}
\newcommand{\cL}{\cal L}
\newcommand{\D}{\cal D}
\newcommand{\Dhalf}{{D\over 2}}
\def\eps{\epsilon}
\def\epshalf{{\epsilon\over 2}}
\def\lag{( -\partial^2 + V)}
\def\freeexp{{\rm e}^{-\int_0^Td\tau {1\over 4}\dot x^2}}
\def\kinb{{1\over 4}\dot x^2}
\def\kinf{{1\over 2}\psi\dot\psi}
\def\expk{{\rm exp}\biggl[\,\sum_{i<j=1}^4 G_{Bij}k_i\cdot k_j\biggr]}
\def\expp{{\rm exp}\biggl[\,\sum_{i<j=1}^4 G_{Bij}p_i\cdot p_j\biggr]}
\def\expshort{{\e}^{\half G_{Bij}k_i\cdot k_j}}
\def\expabb{{\e}^{(\cdot )}}
\def\epseps#1#2{\varepsilon_{#1}\cdot \varepsilon_{#2}}
\def\epsk#1#2{\varepsilon_{#1}\cdot k_{#2}}
\def\kk#1#2{k_{#1}\cdot k_{#2}}
\def\G#1#2{G_{B#1#2}}
\def\Gp#1#2{{\dot G_{B#1#2}}}
\def\GF#1#2{G_{F#1#2}}
\def\Dab{{(x_a-x_b)}}
\def\Dsq{{({(x_a-x_b)}^2)}}
\def\PITD{{(4\pi T)}^{-{D\over 2}}}
\def\4piTD{{(4\pi T)}^{-{D\over 2}}}
\def\4piT4{{(4\pi T)}^{-2}}
\def\TintmD{{\dps\int_{0}^{\infty}}{dT\over T}\,e^{-m^2T}
    {(4\pi T)}^{-{D\over 2}}}
\def\Tintm4{{\dps\int_{0}^{\infty}}{dT\over T}\,e^{-m^2T}
    {(4\pi T)}^{-2}}
\def\Tintm{{\dps\int_{0}^{\infty}}{dT\over T}\,e^{-m^2T}}
\def\Tint{{\dps\int_{0}^{\infty}}{dT\over T}}
\def\np{n_{+}}
\def\nm{n_{-}}
\def\Np{N_{+}}
\def\Nm{N_{-}}
\newcommand{\slG}{{{\dot G}\!\!\!\! \raise.15ex\hbox {/}}}
\newcommand{\Gd}{{\dot G}}
\newcommand{\Gund}{{\underline{\dot G}}}
\newcommand{\Gdd}{{\ddot G}}
\def\GBd12{{\dot G}_{B12}}
\def\Dx{\dps\int{\cal D}x}
\def\Dy{\dps\int{\cal D}y}
\def\Dpsi{\dps\int{\cal D}\psi}
\def\dint#1{\int\!\!\!\!\!\int\limits_{\!\!#1}}
\def\ddtau{{d\over d\tau}}
\def\ie{\hbox{$\textstyle{\int_1}$}}
\def\iz{\hbox{$\textstyle{\int_2}$}}
\def\id{\hbox{$\textstyle{\int_3}$}}
\def\ldop{\hbox{$\lbrace\mskip -4.5mu\mid$}}
\def\rdop{\hbox{$\mid\mskip -4.3mu\rbrace$}}
%
\newcommand{\1}{{\'\i}}
\newcommand{\no}{\noindent}
\def\non{\nonumber}
\def\dps{\displaystyle}
\def\sy{\scriptscriptstyle}
\def\sy{\scriptscriptstyle}

%

\newcommand{\bea}{\begin{eqnarray}}  
\newcommand{\eea}{\end{eqnarray}}  
\def\eqa{&=&}  
\def\ccr{\nonumber\\}  
  
\def\a{\alpha}
\def\b{\beta}
\def\m{\mu}
\def\n{\nu}
\def\r{\rho}
\def\s{\sigma}
\def\ep{\epsilon}

\def\cosech{\rm cosech}
\def\sech{\rm sech}
\def\coth{\rm coth}
\def\tanh{\rm tanh}

\def\sqr#1#2{{\vcenter{\vbox{\hrule height.#2pt  
     \hbox{\vrule width.#2pt height#1pt \kern#1pt  
           \vrule width.#2pt}  
       \hrule height.#2pt}}}}  
\def\square{\mathchoice\sqr66\sqr66\sqr{2.1}3\sqr{1.5}3}  
  
\def\appendix{\par\clearpage
  \setcounter{section}{0}
  \setcounter{subsection}{0}
  \def\@sectname{Appendix~}
  \def\theequation{\thesection\arabic{equation}}
  \def\thesection{\Alph{section}}}
 
\def\thefigures#1{\par\clearpage\section*{Figures\@mkboth
  {FIGURES}{FIGURES}}\list
  {Fig.~\arabic{enumi}.}{\labelwidth\parindent\advance
\labelwidth -\labelsep
      \leftmargin\parindent\usecounter{enumi}}}
\def\figitem#1{\item\label{#1}}
\let\endthefigures=\endlist
 
\def\thetables#1{\par\clearpage\section*{Tables\@mkboth
  {TABLES}{TABLES}}\list
  {Table~\Roman{enumi}.}{\labelwidth-\labelsep
      \leftmargin0pt\usecounter{enumi}}}
\def\tableitem#1{\item\label{#1}}
\let\endthetables=\endlist
 
\def\@sect#1#2#3#4#5#6[#7]#8{\ifnum #2>\c@secnumdepth
     \def\@svsec{}\else
     \refstepcounter{#1}\edef\@svsec{\@sectname\csname the#1\endcsname
.\hskip 1em }\fi
     \@tempskipa #5\relax
      \ifdim \@tempskipa>\z@
        \begingroup #6\relax
          \@hangfrom{\hskip #3\relax\@svsec}{\interlinepenalty \@M #8\par}
        \endgroup
       \csname #1mark\endcsname{#7}\addcontentsline
         {toc}{#1}{\ifnum #2>\c@secnumdepth \else
                      \protect\numberline{\csname the#1\endcsname}\fi
                    #7}\else
        \def\@svse=chd{#6\hskip #3\@svsec #8\csname #1mark\endcsname
                      {#7}\addcontentsline
                           {toc}{#1}{\ifnum #2>\c@secnumdepth \else
                             \protect\numberline{\csname the#1\endcsname}\fi
                       #7}}\fi
     \@xsect{#5}}
 
\def\@sectname{}
%
%
\def\eg{\hbox{\it e.g.}}        \def\cf{\hbox{\it cf.}}
\def\etal{\hbox{\it et al.}}
\def\dash{\hbox{---}}
\def\bR{\mathop{\bf R}}
\def\bC{\mathop{\bf C}}
\def\eq#1{{eq. \ref{#1}}}
\def\eqs#1#2{{eqs. \ref{#1}--\ref{#2}}}
\def\lie{\hbox{\it \$}} 
\def\partder#1#2{{\partial #1\over\partial #2}}
\def\secder#1#2#3{{\partial^2 #1\over\partial #2 \partial #3}}
\def\abs#1{\left| #1\right|}
\def\ltap{\ \raisebox{-.4ex}{\rlap{$\sim$}} \raisebox{.4ex}{$<$}\ }
\def\gtap{\ \raisebox{-.4ex}{\rlap{$\sim$}} \raisebox{.4ex}{$>$}\ }
\def\contract{\makebox[1.2em][c]{
        \mbox{\rule{.6em}{.01truein}\rule{.01truein}{.6em}}}}
%
\def\com#1#2{
        \left[#1, #2\right]}
%
%
\def\bentarrow{\:\raisebox{1.3ex}{\rlap{$\vert$}}\!\rightarrow}
\def\longbent{\:\raisebox{3.5ex}{\rlap{$\vert$}}\raisebox{1.3ex}%
        {\rlap{$\vert$}}\!\rightarrow}
\def\onedk#1#2{
        \begin{equation}
        \begin{array}{l}
         #1 \\
         \bentarrow #2
        \end{array}
        \end{equation}
                }
\def\dk#1#2#3{
        \begin{equation}
        \begin{array}{r c l}
        #1 & \rightarrow & #2 \\
         & & \bentarrow #3
        \end{array}
        \end{equation}
                }
\def\dkp#1#2#3#4{
        \begin{equation}
        \begin{array}{r c l}
        #1 & \rightarrow & #2#3 \\
         & & \phantom{\; #2}\bentarrow #4
        \end{array}
        \end{equation}
                }
\def\bothdk#1#2#3#4#5{
        \begin{equation}
        \begin{array}{r c l}
        #1 & \rightarrow & #2#3 \\
         & & \:\raisebox{1.3ex}{\rlap{$\vert$}}\raisebox{-0.5ex}{$\vert$}%
        \phantom{#2}\!\bentarrow #4 \\
         & & \bentarrow #5
        \end{array}
        \end{equation}
                }
\newcommand{\nc}{\newcommand}
\nc{\spa}[3]{\left\langle#1\,#3\right\rangle}
\nc{\spb}[3]{\left[#1\,#3\right]}
\nc{\ksl}{\not{\hbox{\kern-2.3pt $k$}}}
\nc{\hf}{\textstyle{1\over2}}
\nc{\pol}{\varepsilon}
\nc{\tq}{{\tilde q}}
\nc{\esl}{\not{\hbox{\kern-2.3pt $\pol$}}}
\renewcommand{\theequation}{\arabic{section}.\arabic{equation}}
\renewcommand{\arraystretch}{2.5}
\def\R{1\!\!{\rm R}}
\def\Eins{\mathord{1\hskip -1.5pt
\vrule width .5pt height 7.75pt depth -.2pt \hskip -1.2pt
\vrule width 2.5pt height .3pt depth -.05pt \hskip 1.5pt}}
\newcommand{\symb}{\mbox{symb}}
\renewcommand{\arraystretch}{2.5}
\def\GBd12{{\dot G}_{B12}}
\def\mneg{\!\!\!\!\!\!\!\!\!\!}
\def\Mneg{\!\!\!\!\!\!\!\!\!\!\!\!\!\!\!\!\!\!\!\!}
\def\non{\nonumber}
\def\beqn*{\begin{eqnarray*}}
\def\eqn*{\end{eqnarray*}}
\def\sy{\scriptscriptstyle}
\def\footstrut{\baselineskip 12pt}
\def\square{\kern1pt\vbox{\hrule height 1.2pt\hbox{\vrule width 1.2pt
   \hskip 3pt\vbox{\vskip 6pt}\hskip 3pt\vrule width 0.6pt}
   \hrule height 0.6pt}\kern1pt}
\def\np{n_{+}}
\def\nm{n_{-}}
\def\Np{N_{+}}
\def\Nm{N_{-}}
\def\exmn{\Bigl(\mu \leftrightarrow \nu \Bigr)}
\def\slash#1{#1\!\!\!\raise.15ex\hbox {/}}
\def\dint#1{\int\!\!\!\!\!\int\limits_{\!\!#1}}
\def\bra#1{\langle #1 |}
\def\ket#1{| #1 \rangle}
\def\vev#1{\langle #1 \rangle}
\def\rightvac{\mid 0\rangle}
\def\leftvac{\langle 0\mid}
\def\dps{\displaystyle}
\def\sy{\scriptscriptstyle}
\def\half{{1\over 2}}
\def\third{{1\over3}}
\def\fourth{{1\over4}}
\def\fifth{{1\over5}}
\def\sixth{{1\over6}}
\def\seventh{{1\over7}}
\def\eigth{{1\over8}}
\def\ninth{{1\over9}}
\def\tenth{{1\over10}}
\def\pa{\partial}
\def\ddtau{{d\over d\tau}}
\def\ge{\hbox{\textfont1=\tame $\gamma_1$}}
\def\gz{\hbox{\textfont1=\tame $\gamma_2$}}
\def\gd{\hbox{\textfont1=\tame $\gamma_3$}}
\def\go{\hbox{\textfont1=\tamt $\gamma_1$}}
\def\gt{\hbox{\textfont1=\tamt $\gamma_2$}}
\def\gth{\hbox{\textfont1=\tamt $\gamma_3$}} 
\def\gf{\hbox{$\gamma_5\;$}}
\def\ie{\hbox{$\textstyle{\int_1}$}}
\def\iz{\hbox{$\textstyle{\int_2}$}}
\def\id{\hbox{$\textstyle{\int_3}$}}
\def\ldop{\hbox{$\lbrace\mskip -4.5mu\mid$}}
\def\rdop{\hbox{$\mid\mskip -4.3mu\rbrace$}}
\def\eps{\epsilon}
\def\epshalf{{\epsilon\over 2}}
\def\e{\mbox{e}}
\def\mn{{\mu\nu}}
\def\exmn{{(\mu\leftrightarrow\nu )}}
\def\ab{{\alpha\beta}}
\def\exab{{(\alpha\leftrightarrow\beta )}}
\def\g{\mbox{g}}
\def\kinb{{1\over 4}\dot x^2}
\def\kinf{{1\over 2}\psi\dot\psi}
\def\expk{{\rm exp}\biggl[\,\sum_{i<j=1}^4 G_{Bij}k_i\cdot k_j\biggr]}
\def\expp{{\rm exp}\biggl[\,\sum_{i<j=1}^4 G_{Bij}p_i\cdot p_j\biggr]}
\def\expshort{{\e}^{\half G_{Bij}k_i\cdot k_j}}
\def\expabb{{\e}^{(\cdot )}}
\def\epseps#1#2{\varepsilon_{#1}\cdot \varepsilon_{#2}}
\def\epsk#1#2{\varepsilon_{#1}\cdot k_{#2}}
\def\kk#1#2{k_{#1}\cdot k_{#2}}
\def\G#1#2{G_{B#1#2}}
\def\Gp#1#2{{\dot G_{B#1#2}}}
\def\GF#1#2{G_{F#1#2}}
\def\Dab{{(x_a-x_b)}}
\def\Dsq{{({(x_a-x_b)}^2)}}
\def\lag{( -\partial^2 + V)}
\def\PITD{{(4\pi T)}^{-{D\over 2}}}
\def\4piTD{{(4\pi T)}^{-{D\over 2}}}
\def\4piT4{{(4\pi T)}^{-2}}
\def\TintmD{{\dps\int_{0}^{\infty}}{dT\over T}\,e^{-m^2T}
    {(4\pi T)}^{-{D\over 2}}}
\def\Tintm4{{\dps\int_{0}^{\infty}}{dT\over T}\,e^{-m^2T}
    {(4\pi T)}^{-2}}
\def\Tintm{{\dps\int_{0}^{\infty}}{dT\over T}\,e^{-m^2T}}
\def\Tint{{\dps\int_{0}^{\infty}}{dT\over T}}
\def\pint{{\dps\int}{dp_i\over {(2\pi)}^d}}
\def\Dx{\dps\int{\cal D}x}
\def\Dy{\dps\int{\cal D}y}
\def\Dpsi{\dps\int{\cal D}\psi}
\def\Tr{{\rm Tr}\,}
\def\tr{{\rm tr}\,}
\def\sumij{\sum_{i<j}}
\def\freeexp{{\rm e}^{-\int_0^Td\tau {1\over 4}\dot x^2}}
\def\arraystretch{2.5}
\def\Ge{\mbox{GeV}}
\def\dA{\partial^2}
\def\DA{\sqsubset\!\!\!\!\sqsupset}
\def\FFdual{F\cdot\tilde F}
\def\mn{{\mu\nu}}
\def\rs{{\rho\sigma}}
\def\oplusotimes{{{\lower 15pt\hbox{$\scriptscriptstyle \oplus$}}\atop{\otimes}}}
\def\perppar{{{\lower 15pt\hbox{$\scriptscriptstyle \perp$}}\atop{\parallel}}}
\def\oopp{{{\lower 15pt\hbox{$\scriptscriptstyle \oplus$}}\atop{\otimes}}\!{{\lower 15pt\hbox{$\scriptscriptstyle \perp$}}\atop{\parallel}}}
%
%
\def\bbbr{{\rm I\!R}}
\def\bbbone{{\mathchoice {\rm 1\mskip-4mu l} {\rm 1\mskip-4mu l}
{\rm 1\mskip-4.5mu l} {\rm 1\mskip-5mu l}}}
\def\bbbz{{\mathchoice {\hbox{$\sf\textstyle Z\kern-0.4em Z$}}
{\hbox{$\sf\textstyle Z\kern-0.4em Z$}}
{\hbox{$\sf\scriptstyle Z\kern-0.3em Z$}}
{\hbox{$\sf\scriptscriptstyle Z\kern-0.2em Z$}}}}

\renewcommand{\thefootnote}{\protect\arabic{footnote}}
%

\begin{center}
{\huge\bf }
\vspace{5pt}

{\huge\bf Photonic processes in Born-Infeld theory}
\vskip1.3cm

{\large Jos\'e Manuel D\'avila$^{a}$, 
Christian Schubert$^{b}$, Mar\'ia Anabel Trejo$^{b}$}
\\[1.5ex]

\begin{itemize}
\item [$^a$]
{\it 
Facultad de Ciencias, Universidad Aut\'onoma del Estado de M\'exico,\\
Instituto Literario 100, C.P. 50000, Toluca, M\'exico.\\
}
\item [$^b$]
{\it 
Instituto de F\'{\i}sica y Matem\'aticas
\\
Universidad Michoacana de San Nicol\'as de Hidalgo\\
Edificio C-3, Apdo. Postal 2-82\\
C.P. 58040, Morelia, Michoac\'an, M\'exico.\\
}
\end{itemize}
\end{center}
\vspace{1cm}
 {\large \bf Abstract:}
\begin{quotation}
We study the processes of photon-photon scattering and
photon splitting in a magnetic field in Born-Infeld theory. 
In both cases we combine the terms from the tree-level Born-Infeld
Lagrangian with the usual one-loop QED contributions, where those are
approximated by the Euler-Heisenberg Lagrangian, including also the interference terms. 
For photon-photon scattering we obtain the total cross section in the low-energy
approximation. For photon splitting we compute the total absorption coefficient in
the hexagon (weak field) approximation, and also show that, due to the non-birefringence
property of Born-Infeld theory, the selection rules found by Adler for the QED case
continue to hold in this more general setting. 
We discuss the bounds on the free parameter of Born-Infeld theory
that may be obtained from this type of processes.
 
\end{quotation}
\vfill\eject
\pagestyle{plain}
\setcounter{page}{1}
\setcounter{footnote}{0}

\vspace{10pt}
\section{Introduction}
\label{intro}
\renewcommand{\theequation}{1.\arabic{equation}}
\setcounter{equation}{0}

In their famous 1934 paper \cite{borinf}, Born and Infeld proposed the following 
Lagrangian as a nonlinear generalization of electrodynamics:

\begin{equation}
 \label{BILag}
\mathcal{L}^{BI}
=-b^{2}\sqrt{1-\frac{2s}{b^{2}} -\frac{p^{2}}{b^{4}}}+b^{2}.
\non
\end{equation}
Here $s$ and $p$ are the two invariants of the Maxwell field,
 
\begin{eqnarray}
 \label{ec:ese1}
s\equiv -\frac{1}{4}F^{\mu\nu}F_{\mu\nu} &=& \frac{1}{2}(E^{2}-B^{2}),\\
\label{ec:pe1}
p\equiv -\frac{1}{4}\tilde{F}^{\mu\nu}F_{\mu\nu} &=& \vec{E}\cdotp\vec{B},
\end{eqnarray}
($p$ is only a pseudo-invariant and thus must appear squared.)

Alternatively, the Born-Infeld Lagrangian (`BIL') can also be written in determinantal form,

\begin{equation}
 \label{BILagdet}
\mathcal{L}^{BI}=-b^{2}\sqrt{-{\rm det}\left( \eta_{\mu\nu} + \frac{1}{b}F_{\mu\nu} \right)} + b^{2}\sqrt{-{\rm det}(\eta_{\mu\nu})},
\end{equation}
where $F$ is the field strength tensor. 

In the limit of large $b$ the Lagrangian (\ref{BILag}) reduces to the Maxwell Lagrangian $\mathcal{L}^{M} = s= \frac{1}{2}(E^2-B^2)$, as can be
seen by expanding $\mathcal{L}^{BI}$ in powers of $1/b$:

\begin{equation}
 \label{BILagexpand}
\mathcal{L}^{BI}=s+\frac{1}{2b^2}\left( s^2+p^2 \right)+\frac{1}{2b^4}\left( s^3+sp^2 \right) + \ldots .
\end{equation}
The higher order terms correspond to quartic, sextic, etc. photon vertices. 

Originally Born and Infeld attempted to fix $b$ by equating the electromagnetic self energy of the
electron, which is finite in their theory, with its mass energy. This leads to the following numerical
value of $b$,

\begin{equation}
 \label{fixb}
b =1.2\times 10^{20}\frac{V}{m} \, .
\label{defb}
\end{equation}
Nowadays such an interpretation is hardly viable (see, e.g., \cite{soragr}), but 
Born--Infeld theory, with $b$ as a free parameter,
is still considered the prototypical example of a nonlinear generalization of
electrodynamics. This is not only because of its concise determinant formulation (\ref{BILagdet}),
but also because it shares with Maxwell theory the important property 
of not leading to birefringence; that is, the velocity of photon propagation, although dependent
on the frequency, does not depend on the photon polarization. This property is not shared by
more general nonlinear theories of electromagnetism.

Although Born and Infeld thought of their theory as a substitute for Maxwell theory at the fundamental level,
as a quantum field theory it is not renormalizable, so that nowadays it seems more natural to think of
the Born--Infeld Lagrangian (`BIL') as an effective one. And indeed, in 1985 the Born--Infeld theory
acquired new relevance through the discovery by Fradkin and Tseytlin \cite{fratse} that the same
Lagrangian appears also as a low-energy effective Lagrangian in open string theory
(on the other hand, it has been shown \cite{hagiwara} that no combination of scalar, spinor, and vector particles
alone can, assuming the standard couplings to an abelian gauge field, generate the BIL as a 
one-loop effective Lagrangian). In this context there are also derivative corrections to the BIL, some of
which have been computed \cite{rooeen,wyllard}.

More recently, the BIL has become also an important ingredient for brane theories (see, e.g., \cite{zwiebach-book},
which contains also an excellent introduction to Born--Infeld theory and nonlinear electrodynamics). 
In this context the determinantal definition (\ref{BILagdet}) is very convenient, since it admits an
immediate generalization to other (even) dimensions. Moreover, the BIL can appear both as an effective Lagrangian
or as a fundamental one, now motivated by T-duality.

Despite of the present ubiquitous appearance of the Born--Infeld Lagrangian in field theory, relatively little effort has been devoted to
taking it seriously as an alternative to Maxwell electrodynamics. In the absence of experimental evidence for deviations from
Maxwell theory, here the goal must be to establish successively stricter lower bounds on the parameter $b$.
Since this parameter also represents the maximal possible field strength in Born--Infeld theory, it seems logical 
in this context to focus on strong-field atomic physics \cite{rafugr71,rafugr72,soragr,iaczav,carkie,fragar}. 
And indeed, as far as is known to the authors the strongest bound
on $b$ presently available is the one obtained by Soff, Rafelski and Greiner \cite{soragr} from muonic transitions in lead,


\bear
b \geq 1.7 \times 10^{22} \frac{V}{m} \, . 
\label{soffbound}
\ear
This is already considerably  beyond the original Born-Infeld value (\ref{defb}). 
However, the mixture of nonlinear electrodynamics, strong-field physics and
quantum mechanics involved in this type of estimate is a subtle one, and it is natural to ask
what bounds on $b$ can be obtained by purely photonic processes not involving the
electrostatic potential between point charges. 
In this paper we will study two such processes
that are the most obvious ones for testing the four-point and six-point vertices contained in
the weak-field expansion of the Born--Infeld Lagrangian (\ref{BILagexpand}), namely photon-photon scattering and
photon splitting in a magnetic field. 

In the context of such purely photonic processes the classical Born--Infeld Lagrangian
enters naturally  in competition with the quantum Euler-Heisenberg Lagrangian (`EHL'), 
the one-loop low-energy effective Lagrangian of QED \cite{eulhei} in the constant field approximation.
We recall the standard proper-time representation of this Lagrangian (see, e.g., \cite{ditreu-book,geraldrev})

\bear
\mathcal {L}^{EH}&=& - {1\over 8\pi^2}
\int_0^{\infty}{dT\over T^3}
\,\e^{-m^2T}
\biggl\lbrack
{(eaT)(ebT)\over {\rm tanh}(eaT)\tan(ebT)} 
\nonumber\\&&\hspace{70pt}
- {e^2\over 3}(a^2-b^2)T^2 -1
\biggr\rbrack \, .
\label{ehspin}
\ear
Here $T$ is the proper-time of the loop fermion, $m$ its mass, and $a,b$ are related to the
invariants $s$ and $p$ by $b^2-a^2 = 2s$ and $ab = p$.
The two subtraction terms implement the renormalization of charge and vacuum
energy. 

\noindent
Thus in this paper we will generally consider a total Lagrangian

\bear
\mathcal {L}^{\rm total} &=& \sum_f\mathcal {L}^{EH}_f + \mathcal {L}^{BI} 
\label{defLtotal}
\ear
that is, adding the EHLs generated by all the charged standard model fermions to the BIL.
For the type of processes considered only the electronic EHL will be phenomenologically relevant, though
(for the same reason we do not bother here to include the analogue of the EHL involving a $W^{\pm}$ in the loop \cite{skavan}).

The EHL contains a wealth of information on photonic processes involving
a single virtual electron-positron pair in the vacuum \cite{ditgie-book,ditgie}. However,  it can be expected to be a good approximation 
only for photon energies $\omega$ much smaller than the electron mass, since otherwise derivative
corrections to the  EHL come into play which carry factors of $\omega/m$. 
If to this constant field (resp. low photon energy) approximation one adds the approximation of a weak field (resp. low
photon beam intensity), then one can further simplify (\ref{ehspin}) by expanding  out in powers of the field. The first two
terms of this expansion are 

\begin{eqnarray}
\mathcal{L}^{EH}=&&\frac{2\alpha ^{2}}{45m^4}\left(4s^2+7p^2 \right)
+ \frac{32 \pi\alpha^3}{315 m^{8}}\left(8 s^3+ 13sp^2 \right)
+ \ldots \, , 
\non\\
 \label{EHexp}
\end{eqnarray}
where $\alpha = \frac{e^2}{4 \pi}$ is the fine structure constant.

\begin{figure}[h]
\centering
\hspace{20pt}\includegraphics[scale=.8]{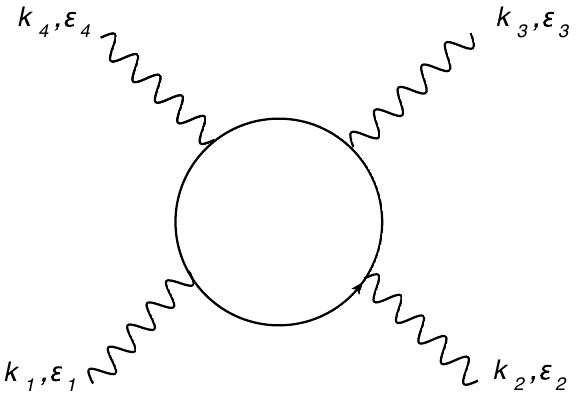}
\hfill
\includegraphics[scale=.8]{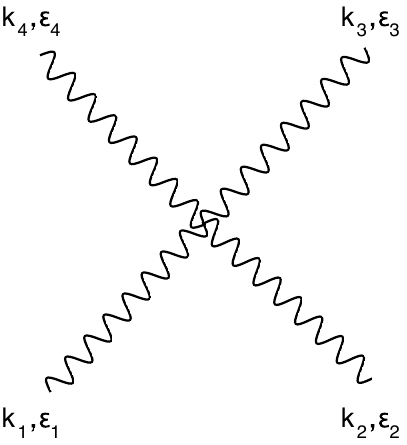}
\hspace{20pt}
\caption{Photon-photon scattering in QED (left) and BI theory (right).}
\label{fig4point}
\end{figure}

Here the first term holds the information on the one-loop photon-photon scattering amplitude in the low energy limit, see fig. \ref{fig4point}, left.

\begin{figure}[h]
\centering
\hspace{40pt}\includegraphics[scale=.7]{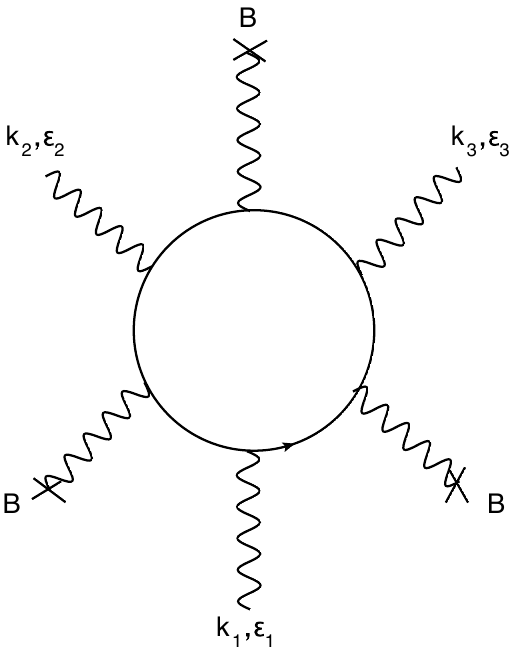}
\hspace{60pt}
\caption{Photon splitting in a magnetic field in QED.}
\label{fighexagonQED}
\end{figure}

Here and in the following a diagram is understood to include also all relevant permutations of the external legs.
The second term in (\ref{EHexp}) corresponds to the low energy limit of various photonic processes, of which the most important one is 
the splitting of one photon into two in a magnetic field \cite{biabia,adler71}. In vacuum this process
has, despite of its very special kinematics - energy-momentum conservation forces all three photons to be collinear - 
a non vanishing two-body phase space, but the matrix element vanishes on account of Furry's theorem.
In an external field it becomes possible, since the odd number of photons can
be balanced by an odd number of interactions with the field. For a generic magnetic field, the lowest order contribution to the 
matrix element would
again be given by the photon-photon scattering diagram (fig. \ref{fig4point}, left), now with one of the photons replaced by an interaction with
the field. In the case of a constant magnetic field, the most important one for phenomenology, it turns out that
this box diagram contribution to the matrix element vanishes due to the collinear kinematics; 
therefore the leading contribution is presented by the hexagon diagram,  shown in fig. \ref{fighexagonQED}.

\no
In this diagram the crosses denote the interactions with the magnetic field.
This contribution is non-vanishing, but still has the
very special property of being exactly given by its low-energy (small $\omega/m$) limit. It thus can be exactly computed
from the Euler-Heisenberg Lagrangian, which was done by
Z. and I. Bialynicka-Birula \cite{biabia}. A full calculation of the photon splitting process, summing
all diagrams with an arbitrary (odd) number of interactions with the field, requires more advanced methods, and was
first achieved by Adler \cite{adler71} (see also \cite{bamish,17,ditgie-book}). 

A full treatment of the photon splitting process must, however, also take into account the fact 
that the presence of the magnetic field will modify the photon dispersion relations. The vacuum
will acquire a nontrivial index of refraction, which moreover turns out to depend on the photon
polarization, leading to birefringence \cite{toll,baibre,adler71}. The effect of this on the photon splitting process is
that, depending on the chosen combination of photon polarizations, either the collinearity of the
photons is slightly modified, or else the process becomes impossible altogether. As shown in \cite{adler71},
combining this dispersion-induced selection rule with CP invariance leaves, up to order $\alpha$ corrections,
the splitting of a perpendicularly polarized photon into two parallely polarized ones as the only 
possible polarization choice\footnote{Here by $\perp$ ($\parallel$) we denote a polarization vector which is perpendicular (parallel)
to the plane spanned by the photon momentum and the direction of the magnetic field (this convention is opposite
to the one used in \cite{adler71}, but agrees with \cite{ditgie-book}).}. 
Photon splitting thus naturally has a polarizing effect.

Physically the meanings of the BIL and the EHL are worlds apart: the latter describes the effect
of the electron, or other standard model fermions, while the former is (apart from its leading Maxwell
term) associated with new physics. 
Nevertheless, as seen by a comparison of (\ref{BILagexpand}) and (\ref{EHexp}) their weak field expansions 
differ, apart from the replacement of the electron mass $m$ by $\sqrt{b}$ as the intrinsic mass scale, 
only by a change of the numerical coefficients (as a historical curiosity, let us mention that Infeld \cite{infeld}
attempted to determine the fine structure constant by equating ${\cal L}_4^{EH}$ and ${\cal L}_4^{BI}$,
and with $b$ as given in (\ref{defb}) obtained its correct order of magnitude). 
Therefore any physical process encoded in the EHL is expected to have an analogous contribution 
induced by the BIL, and can potentially be used for obtaining a bound on $b$. 
Constraining the Born-Infeld parameter through low-energy photonic processes provides not only the most
direct possible way of testing the presence of the nonlinear photon vertices, but has also the advantage
that the obtained  bounds will hold irrespectively of whether the BIL is considered as fundamental or effective;
derivative corrections, which would constitute the difference between the both scenarios, 
would come with factors of $\omega^2/b$
that will be negligible for the values of $b$ that are presently still viable. 

Photon amplitudes in Born-Infeld theory have already been discussed by many authors, both from
a theoretical and a phenomenological perspective. On the theory side, there has been interest in the fact that
the $N$-photon helicity amplitudes in Born-Infeld theory are, at least at the tree-level, helicity
conserving. This was shown in \cite{rossel} to be a consequence of the invariance under the
$U(1)$ duality rotation of $F$ and $\tilde F$. 
When viewing the BIL as induced by open string theory
this mechanism relates to S-duality \cite{garousi}. 
Phenomenologically, the recent interest in Born-Infeld theory, and more generally nonlinear electrodynamics,
is due to the construction of high-power laser facilities such as POLARIS, HERCULES, VIRGO 
and ELI, which will allow one to test QED in hitherto completely unexplored sectors.

Despite of all this recent activity, to the best of our knowledge the cross section for photon-photon scattering in BI theory
is not available in the literature. 
In section \ref{scatt} we will calculate this cross section, combining the standard QED contribution 
(Fig. \ref{fig4point}, left)  with the one of the four-photon vertex of Born-Infeld theory (Fig. \ref{fig4point}, right), 
and including also the interference term between them.

\begin{figure}[h]
\centering
\hspace{120pt}\includegraphics[scale=.7]{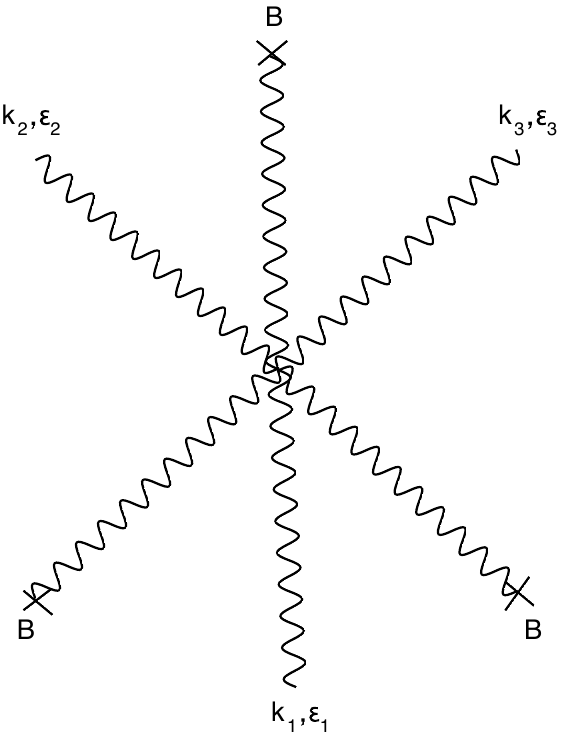}
\hfill
\hspace{60pt}
\caption{Diagram for photon splitting in BI theory.}
\label{fighexagonBI}
\end{figure}

In section \ref{split} we analogously study the photon splitting 
process in a constant magnetic field for Born-Infeld theory, combining the 
QED contribution with the one induced by the sextic Born-Infeld vertex, Fig. \ref{fighexagonBI}. 

We also show that the six-point diagrams with two quartic vertices can be omitted,
and that in BI theory the same selection rules hold for this process as in QED. 
In the photon splitting case, we are not aware
of any previous discussions for Born-Infeld theory. 

In section \ref{exp} we discuss the use of these results for obtaining
lower bounds on $b$ from laser physics respectively neutron star physics. 
We summarize our findings in section \ref{conclusions}. 

\section{Photon-photon scattering}
\label{scatt}
\renewcommand{\theequation}{2.\arabic{equation}}
\setcounter{equation}{0}

In BI theory, photon-photon scattering occurs already at the tree-level, due to the quartic vertex
implied by the first nontrivial term on the right hand side of (\ref{BILagexpand}) and depicted 
on the right in Fig. \ref{fig4point}. The calculation of the cross section from this vertex 
is analogous to the one of the photon-photon cross section in QED in the low-energy limit from
the quartic term of the EHL. We will thus first retrace this textbook calculation, following \cite{itzzub-book}.

\subsection{Photon-photon scattering in QED}
\label{scattQED}

Let us denote the quartic term in (\ref{EHexp}) by ${\cal L}_4^{EH}$, and rewrite it using
the identity 

\begin{eqnarray}
 \bigl( F\cdotp \tilde{F} \bigr)^{2} &=& 4\tr(F^{4})-2(\tr(F^{2}))^{2}.
 \label{id1}
\end{eqnarray}
It then reads

\begin{eqnarray}
{\cal L}_4^{EH} &=& \frac{2\alpha^2}{45 m^4} \Bigl\lbrack \frac{7}{4}(\tr(F^4)-\frac{5}{8} (\tr(F^2))^2\Bigr\rbrack
\, .
\label{LEH4}
\end{eqnarray} 
We introduce for each photon leg its field strength tensor,

\begin{eqnarray}
F_{i\mu \nu} \equiv (k_{i\mu}\,\varepsilon_{i\nu}-k_{i\nu}\,\varepsilon_{i\mu}) \, .
\label{Fi}
\end{eqnarray}
We replace in ${\cal L}_4^{EH}$ each $F$ by the sum
$F_1+F_2+F_3+F_4$, and keep only the terms containing each
$F_1,\ldots,F_4$. 
This gives the photon scattering amplitude $\mathcal{M}$ as
\footnote{In \cite{itzzub-book} it is stated that one should also divide by a combinatorial factor
of $4!$, however this is erroneous, as was confirmed to us by J.B. Zuber.}:

\begin{eqnarray}
\mathcal{M}&=&-\frac{2\,\alpha^2}{45\,m^4}\Big[5\big(\mbox{tr}(F_1F_2)\,\mbox{tr}(F_3 F_4)+\mbox{tr}(F_1F_3)\,\mbox{tr}(F_2F_4)
+\mbox{tr}(F_1F_4)\,\mbox{tr}(F_2F_3)\big)\nonumber\\
&&-7\,\mbox{tr}\big(F_1F_2F_3F_4+F_3F_1F_2F_4+F_2F_3F_1F_4+F_3F_2F_1F_4+F_1F_3F_2F_4
\nonumber\\&&
+F_2F_1F_3F_4\big)\Big]\, .\nonumber\\
\label{MQED}
\end{eqnarray}
The unpolarized cross section in the center-of-mass-frame becomes

\begin{eqnarray}
d\sigma&=&\frac{1}{4\,(k_1\cdot k_2)}\int\frac{d^3 k_3}{2\,\omega_3\,(2\,\pi)^3}\frac{d^3 k_4}{2\,\omega_4\,(2\,\pi)^3}(2\,\pi)^4\,
\delta^4(k_3+k_4-k_1-k_2)\,\overline{|\mathcal{M}|^2}\nonumber\\ 
&=&\frac{1}{64\,\omega^2\,(2\,\pi)^2}\,\overline{|\mathcal{M}|^2}\,d\Omega\,,
\end{eqnarray}
where

\begin{eqnarray}
\overline{|\mathcal{M}|^2}\equiv \frac{1}{4}\sum_{\varepsilon_i}|\mathcal{M}|^2\,.
\end{eqnarray}
The sums over polarization can be done using, for each photon leg, the identity

\begin{eqnarray}
\sum_{\varepsilon}F^{*\ \beta}_\alpha\,F^{\ \ \nu}_\mu=-k^\beta\,k^\nu\,g_{\alpha \mu}+k_\mu\,k^\beta\,g_\alpha{}^\nu
+k_\alpha\,k^\nu\,g_\mu{}^\beta-k_\alpha\,k_\mu\,g^{\beta \nu}\,,
\end{eqnarray}
with the result

\begin{eqnarray}\label{799}
\overline{|\mathcal{M}|^2}
&=&\frac{32 \cdot 139}{(90)^2}\frac{\alpha^4}{m^8}
\,\Big[ (k_1 \cdot k_2)^2\,(k_3 \cdot k_4)^2+(k_1 \cdot k_3)^2\,(k_2 \cdot k_4)^2
\nonumber\\  && \hspace{120pt}
+(k_1 \cdot k_4)^2\,(k_2 \cdot k_3)^2\Big]\,.\nonumber\\
\end{eqnarray}
Using the kinematic relations

\begin{eqnarray}
&&k_1 \cdot k_2=k_3 \cdot k_4 = 2\,\omega^2 \,,\nonumber\\
&&k_1 \cdot k_3=k_2 \cdot k_4 = \omega^2\,(1-\mbox{cos}\,\theta)\,, \nonumber\\
&&k_1 \cdot k_4=k_2 \cdot k_3 = \omega^2\,(1+\mbox{cos}\,\theta)\, ,
\end{eqnarray}
one finds 

\begin{eqnarray}
\frac{d\sigma}{d\Omega}=\frac{1}{(2\,\pi)^2}\frac{139}{(90)^2}\alpha^4\left(\frac{\omega}{m}\right)^6
\frac{1}{m^2}(3+\mbox{cos}^2\,\theta)^2
\end{eqnarray}
for the differential cross section, and

\begin{eqnarray}
\sigma=\frac{1}{2\,\pi}\frac{139}{(90)^2}\left( \frac{56}{5}\right)\,\alpha^4\left(\frac{\omega}{m}\right)^6\frac{1}{m^2}
\label{sigmaQED}
\end{eqnarray}
for the total cross section. 
See \cite{aklapo,akhieser,karneu,cotopi} for the generalization to arbitrary energies,
\cite{56} for the generalization to the $N$ - photon case.

\subsection{Photon-photon scattering in Born-Infeld theory}
\label{scattBI}

The calculation of the cross section for the Born-Infeld case is completely analogous. Using the identity
(\ref{id1}) to eliminate $\tilde F$ from the quartic term of the BIL (\ref{BILagexpand}) gives

\begin{equation}
 \label{BIL4fin}
\mathcal{L}_{4}^{BI} = \frac{1}{32b^{2}}\left[ 4{\rm tr}(F^4)-({\rm tr}(F^2))^{2} \right].
\end{equation}
Proceeding in the same way as before one obtains, after a lengthy calculation, 

\begin{eqnarray}
 \label{Msquared}
\overline{ \vert {\cal M}^{\rm BI} \vert ^{2} }&=& \frac{2}{b^{4}}\left[ (k_{1}\cdotp k_{2})^{2}(k_{3}\cdotp k_{4})^{2}+(k_{1}\cdotp k_{3})^{2}(k_{2}\cdotp k_{4})^{2}+(k_{1}\cdotp k_{4})^{2}(k_{2}\cdotp k_{3})^{2} \right].
\non\\
\end{eqnarray}
The differential cross section in the center-of-mass frame becomes

\bear
\frac{d\sigma}{d\Omega} = \frac{1}{64\omega^{2}(2\pi)^{2}}\overline{ \vert {\cal M}^{\rm BI} \vert^{2} }
=\frac{1}{16(2\pi)^{2}}\left( \frac{\omega}{b} \right)^{4}\omega^{2}\left(3+\cos^{2}\theta \right)^{2} \, .
\label{dsigma}
\ear
Note that the angular dependence is the same as in the QED case. 
For the total cross section one obtains

\begin{equation}
 \label{sigmatotal}
\sigma^{\rm BI} =\frac{1}{2\pi}\frac{7}{10} \frac{\omega^{6}}{b^{4}}.
\end{equation}

When contemplating Born-Infeld theory as a phenomenologically viable extension of QED,  
we must include also the standard coupling to fermions. Thus both of the diagrams of Fig. 
\ref{fig4point} must be taken into account, including the interference terms. The total photon-photon
scattering differential and total cross sections become
\begin{eqnarray}
\frac{d\sigma}{d\Omega}=\left( \frac{1}{64\ b^4}+\frac{11\ \alpha^2}{720\ b^2\ m^4}+\frac{139\ \alpha^4}{32400\ m^8}\right)\frac{\omega^6}{\pi^2} \Big(3+\cos^2 \theta\Big)^2,
\end{eqnarray}
and

\begin{eqnarray}
\sigma=\left(\frac{7}{20\ b^4}+\frac{77\ \alpha^2}{225\ b^2\ m^4}+\frac{973\ \alpha^4}{10125\ m^8}\right)\frac{\omega^6}{\pi}.
\label{sigmafin}
\end{eqnarray}

\section{Photon splitting}
\label{split}
\renewcommand{\theequation}{3.\arabic{equation}}
\setcounter{equation}{0}

We proceed to the calculation of the photon splitting amplitude in a constant magnetic field in Born-Infeld theory,
in the leading order in the weak-field expansion. It is easily checked that, for the same kinematical reasons as in
the QED case, also in Born-Infeld theory there is no contribution to the photon splitting amplitude from the quartic
vertex in the BIL (\ref{BILagexpand}). Thus, apart from the same diagram as in QED, Fig. \ref{fighexagonQED},
we have to consider the basic sextic vertex, Fig. \ref{fighexagonBI}.
As in the photon-photon scattering case, we will first retrace the calculation of the standard QED contribution,
and then indicate the necessary modifications to get the Born-Infeld contributions to the amplitude. 

\subsection{Photon splitting in QED}
\label{splitQED}

Since a constant magnetic field cannot absorb four-momentum, the four-vectors $k_1,k_2,k_3$ of the
initial and final photons must obey energy-momentum conservation by themselves,

\bear
\label{ec:conser1}
k_{1}=\omega_{1}(1,\hat{k}_{1})=k_{2}+k_{3}=\omega_{2}(1,\hat{k}_{2})+\omega_{3}(1,\hat{k}_{3}).
\ear
Neglecting at first the modification of the photon dispersion relation due to the presence of the magnetic field,
it is easy to see that, in vacuum, this implies collinearity of the three photons,

\bear
 \label{ec:colineal}
\hat{k_{1}}=\hat{k_{2}}=\hat{k_{3}}\, .
\ear
Thus the photon four-vectors are proportional, and one has the vanishing scalar products

\bear
k_{1}^{2}=k_{2}^{2}=k_{3}^{2}=k_{1}\cdot k_{2}=k_{1}\cdot k_{3}=k_{2}\cdot k_{3}=0 \, .
\label{kinrel}
\ear
It is this lack of nonzero Lorentz invariants that leads to the vanishing of the  box diagram contribution to the photon splitting amplitude,
and to the already mentioned property of the leading hexagon diagram contribution to be given by its low-energy limit.
Thus we can compute this hexagon contribution exactly from the sextic term in the expansion of the EHL (\ref{EHexp}).
Using the identity (\ref{id1}) this term becomes

\begin{equation}
 \label{ec:ehlagrangiano}
\mathcal{L}_{6}^{EH}= 8\frac{\alpha^3\pi}{\ m^{8}}\left[ \frac{13}{1260}{\rm tr}(F^2){\rm tr}(F^4) -\frac{1}{280}({\rm tr}(F^2))^3\right] \, .
\end{equation}
Similarly to the photon-photon scattering case, we can get the matrix element from this by substituting 
$F= F_1+F_2+F_3+F+F+F$, where the $F_i$ are the photon field strength tensors and $F$ is the one
of the external field, and selecting the terms of the form $F_1F_2F_3FFF$. But before starting on this, let us
first establish the polarization selection rules. A priori, there are eight possible combinations of polarizations
in the photon splitting process, whose matrix elements we will denote by

\begin{equation}
 \label{ec:elem}
\mathcal{M}\left[ {\parallel \choose \perp}_{1} \to {\parallel \choose \perp}_{2} + {\parallel \choose \perp}_{3} \right].
\end{equation}
As shown in \cite{adler71}, four of them vanish on account of CP invariance; those are
$\mathcal{M}[(\parallel_{1}) \to (\parallel)_{2}+(\parallel)_{3}]$, $\mathcal{M}[(\parallel)_{1}\to (\perp)_{2}+(\perp)_{3}]$, $\mathcal{M}[(\perp)_{1}\to(\parallel)_{2}+(\perp)_{3}]$, and $\mathcal{M}[(\perp)_{1}\to (\perp)_{2}+(\parallel)_{3}]$.
But, as was discussed already in the introduction, here it is important to take also into account the change of the
photon dispersion relation due to the magnetic field. This change can be described by polarization-dependent
indices of refraction \cite{toll,adler71,ditgie-book}

\begin{figure}[h]
\centering
\includegraphics[scale=.61]{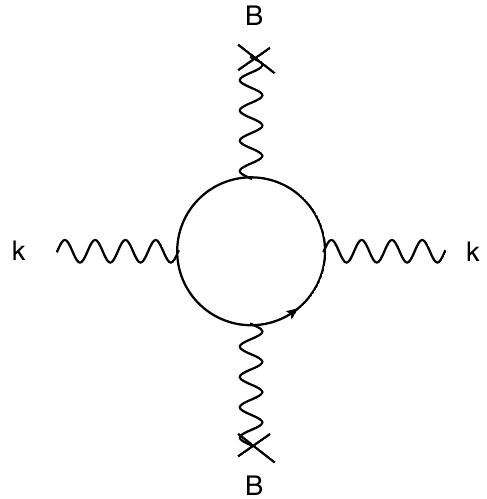}
\caption{Diagram modifying the photon dispersion relation in QED.}
\label{figDispQED}
\end{figure}

\bear
n_{\parallel,\perp}(\omega) &=& \frac{k}{\omega}.
\label{defnpp}
\ear
To lowest order in the field, those are induced by the diagram of Fig. \ref{figDispQED}.

\no
In the low-frequency limit, they are given explicitly by (see, e.g., \cite{ditgie-book})

\bear
n_{\parallel}^{\rm QED} &=& 1 + \frac{14}{45}\frac{\alpha^2}{m^4}B^2\sin^2\theta, \nonumber\\
n_{\perp}^{\rm QED} &=& 1 + \frac{8}{45}\frac{\alpha^2}{m^4}B^2\sin^2\theta. \nonumber\\
\label{nQED}
\ear
Here $\theta$ is the angle between the photon propagation direction and the magnetic field.
It is easy to see \cite{adler71} that, with these nontrivial dispersion relations, photon splitting is compatible
with energy-momentum conservation only if the following condition is fulfilled:

\bear
\Delta \equiv \omega_2n(\omega_2) +\omega_3n(\omega_3)- (\omega_2+\omega_3)n(\omega_2 +\omega_3)
\geq 0
\, .
\label{dispcond}
\ear
In \cite{adler71} it is then shown that, for photon frequencies up to the pair creation threshold $\omega=2m$ and
arbitrary field strengths $B$, of the four polarization choices allowed by CP invariance only the
case $\mathcal{M}[(\perp)_{1}\to (\parallel)_{2}+(\parallel)_{3}]$ fulfills the condition (\ref{dispcond}). 

Returning to the hexagon approximation, choosing polarization vectors corresponding to this
particular component one obtains 
from (\ref{ec:ehlagrangiano}), after a simple calculation (which moreover needs to be done only for
$\sin\theta = 1$, by Lorentz invariance) the matrix element 

\bear
\mathcal{M}[(\perp)_{1}\to(\parallel)_{2}+(\parallel)_{3}] &=& \frac{13}{315}e^3\frac{(eB\sin\theta)^3}{\pi ^2 m^{8}}\omega_1\omega_2\omega_3.
\label{Mps}
\ear
From this the absorption coefficient $\kappa$ is obtained as

\begin{eqnarray}
\label{ec:absorcion3}
\kappa[(\perp)_{1}\to(\parallel)_{2}+(\parallel)_{3}] &=& \frac{1}{32\pi \omega_{1}^2} \int_{0}^{\omega_{1}}
\mathrm{d}\omega_{2}\int_{0}^{\omega_{1}}\mathrm{d}\omega_{3} \delta(\omega_{1}-\omega_{2}-\omega_{3})  \nonumber \\
& & \times \big|\mathcal{M}[(\perp)_{1}\to  (\parallel)_{2}+(\parallel)_{3}]\big|^2  \nonumber\\
&=&\left(\frac{13}{315}\frac{e^3}{\pi ^2 m^{8}}\right)^2\frac{\omega_{1}^{5}(eB\sin\theta)^6}{960\pi}\, .
\end{eqnarray}
For applications it is useful to introduce the ``critical'' field strength $B_{\rm cr} \equiv m^2/e$, and 
rewrite $\kappa$ in terms of dimensionless ratios as 

\bear
\frac{\kappa}{m} = \frac{13^2}{3^5\times 5^3\times 7^2} \frac{\alpha^3}{\pi^2} 
\Bigl(\frac{B}{B_{\rm cr}}\sin\theta \Bigl)^6\Bigl(\frac{\omega_1}{m}\Bigr)^5
\, .
\label{rewritekappa}
\ear
Once the modified dispersion relations are taken into account one needs to also reanalyze the contribution of
the box diagram, as well as the CP-forbidden polarization choices \cite{adler71}. It turns out that now those are in general nonzero, 
but the matrix elements of the former are still down by a factor of order $\alpha$ with respect to the hexagon contribution, 
and the ones of the latter even by a factor of the order of $\alpha (B/B_{\rm cr})^2$.

\subsection{Photon splitting in Born-Infeld theory}
\label{splitBI}

\begin{figure}[h]
\centering
\includegraphics[scale=.61]{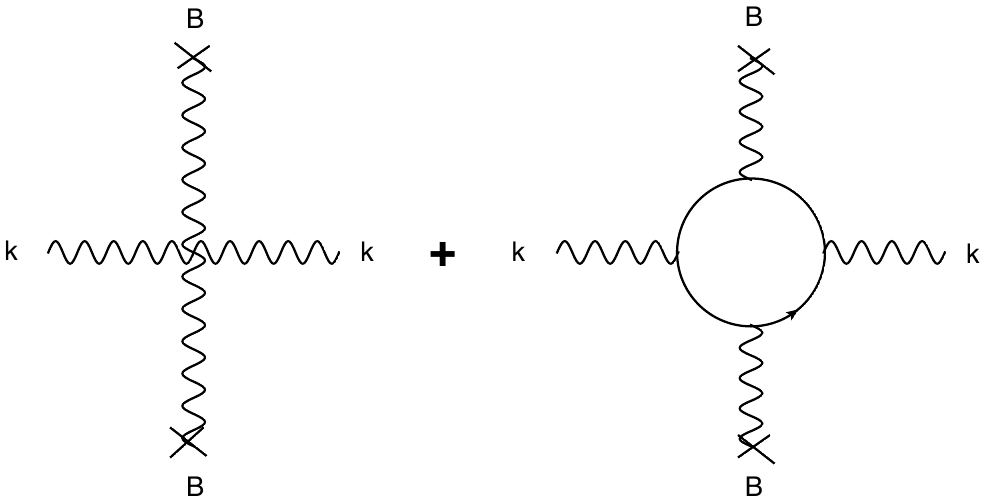}
\caption{Diagrams modifying the photon dispersion relation in BI theory.}
\label{figDispBI}
\end{figure}

Proceeding to the Born-Infeld case, we can see immediately that, at least in the weak field approximation,
there are no essential differences to the QED case. The lowest order contribution to the photon splitting amplitude
now comes from the quartic term in the BIL (\ref{BIL4fin}) with one photon leg substituted by the interaction
with the magnetic field, and with the vacuum dispersion relation it vanishes for the same kinematic reasons
as in QED. The leading contribution to the matrix element thus comes from the sextic vertex, shown in Fig. \ref{fighexagonBI},
and the CP-induced part of the selection rules holds for it as well.
As to the selection rules coming from the modified dispersion relation, these would not hold in pure Born--Infeld theory
due to the absence of birefringence; the photon propagation is modified, but the corresponding index of refraction $n^{\rm BI} (\omega)$
does not depend on polarization, i.e. $n_{\parallel}^{\rm BI} =  n_{\perp}^{\rm BI} = n^{\rm BI}$. 
For example, to lowest order in the field in pure Born--Infeld theory the index of refraction would come from the first of the
diagrams shown in Fig. \ref{figDispBI}, where the quartic vertex now appears with two photon and two magnetic field legs.

\no
It is easily calculated that this diagram alone would yield a refraction index 

\bear
n_{\parallel,\perp}^{\rm BI} = 1 + \half \frac{(B\sin\theta)^2}{b^2}\, ,
\label{nBI}
\ear
confirming the independence of polarization. 
However, if as usual we assume also the presence of the QED diagrams, we will get a total refraction index from the
sum of both diagrams of Fig. \ref{figDispBI} that combines (\ref{nQED}) and (\ref{nBI}) 

 \bear
 n^{\rm total}_{\parallel, \perp} = n^{\rm QED}_{\parallel, \perp} +  \half \frac{(B\sin\theta)^2}{b^2} 
 \label{ntotal}
 \ear
Since a shift of both refraction indices by the same amount drops out of the selection condition (\ref{dispcond})
we then find the same selection rules as above. 
Thus as in the QED case photon splitting is, at leading order in $\alpha$ and in the weak field expansion, possible
only for the combination $(\perp)_{1}\to(\parallel)_{2}+(\parallel)_{3}$. The calculation of the matrix element
from the sextic term in the BIL (\ref{BILagexpand}) again parallels the QED calculation, and we find

\bear
\mathcal{M}^{\rm BI}[(\perp)_{1}\to(\parallel)_{2}+(\parallel)_{3}] = \frac{(B\sin\theta)^3}{b^4} \omega_1\omega_2\omega_3.
\label{MpsBI}
\ear
Adding this to the QED matrix element (\ref{Mps}) leads to the total absorption coefficient

\bear
\kappa^{\rm total}[(\perp)_{1}\to(\parallel)_{2}+(\parallel)_{3}]  =
\biggl\lbrack \frac{1}{b^4} + \frac{13\times 64}{315} \frac{\alpha^3\pi}{ m^8} \biggr\rbrack^2
(B\sin\theta )^6 \frac{\omega_1^5}{960 \pi} \, .
\label{kappatotal}
\ear
As yet another analogy to the QED case, this result for the leading contribution to the photon splitting decay rate
would, for the same kinematic reasons that make the QED hexagon contribution exact in the weak-field limit,
not be affected by the addition of derivative corrections to the sextic vertex in the BIL. Thus, unlike the case of
photon-photon scattering, here even away from the low energy limit $\omega/m \ll 1$ it does not make a difference
whether we consider the BIL as fundamental or effective. 

\begin{figure}[h]
\centering
\hspace{40pt}\includegraphics[scale=.29]{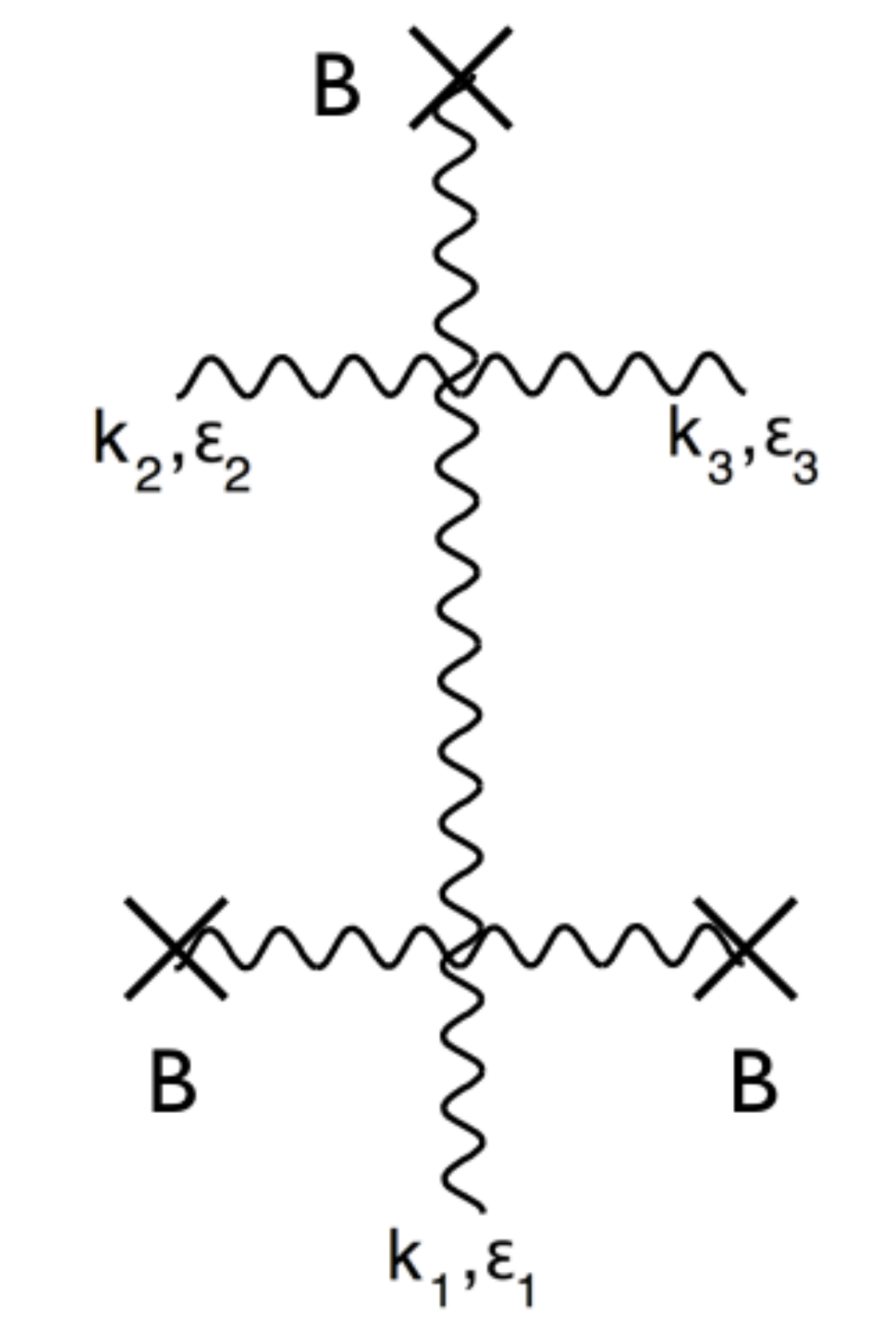}
\caption{Photon splitting in a magnetic field in BI theory.}
\label{figquarticBI}
\end{figure}

It should be noted that, at the same order of the hexagon diagram, in Born-Infeld theory 
there are also diagrams with two quartic vertices connected by a virtual photon, see Fig. \ref{figquarticBI}.

However, for a constant field in this type of graph the momentum of the internal photon is either zero or on-shell,  
so that these contributions will be removed by renormalization (although Born-Infeld theory is not renormalizable, 
the photon propagator must, of course, be renormalized as for the QED case in any halfway realistic setting).

\section{Phenomenological bounds on the Born-Infeld parameter}
\label{exp}
\renewcommand{\theequation}{4.\arabic{equation}}
\setcounter{equation}{0}

Let us now discuss what possible bounds on the Born-Infeld parameter $b$ might be obtained
experimentally through the type of processes which we have discussed here. 

As is well-known, the QED four-photon amplitude is tested in Delbr\"uck scattering,  the scattering
of a photon in the electric field of the atomic nucleus (given by the diagram of fig.1, left, with two
photon legs and two legs representing the interaction with the field), which was discovered 
as early as 1933 and thus even predated the Euler-Heisenberg calculation. However, Delbr\"uck
scattering is of limited relevance for our present investigation, since it is known that for the computation 
of this process the inclusion of derivative corrections to the effective QED Lagrangian is essential. 
Thus our low-energy approximation would not be justified already at the QED level, and for the Born-Infeld part we
would have to assume either the absence of such corrections, or come up with a more specific model.
Even less sense would it make to discuss here processes involving the scattering of high-energy
photons (see the recent \cite{densil} for a discussion of the prospects of a direct verification of photon-photon
scattering at the LHC).  

In our context, the cleanest possible test of the four-vertex would be by laser scattering, where the low-energy
approximation is generally justified. But for optical frequencies the QED cross section (\ref{sigmaQED}) 
is extremely small, due to the factor of $\left(\frac{\omega}{m}\right)^6$. 
The experimental state-of-the-art, presently defined by an experiment performed in 2000
by Bernard et al. \cite{bernard}, still leaves a gap of 18 orders of magnitude between the QED cross section and its
experimental verification. Although with present-day laser technology one could presumably
come significantly closer to the cross section (see \cite{bernard_talk,lundin_etal,tfms,tommic} for various projections of what may be achievable in
laser scattering in the near future), plugging the numbers into (\ref{sigmafin}) this would still not
yield a relevant bound on $b$.  

Alternatively one may also try to test the four-vertex through the low-energy dispersion relations (\ref{nQED}), (\ref{nBI}). 
Here the polarization-independence of the Born-Infeld correction (\ref{nBI})
constitutes a disadvantage, since a measurement of the absolute magnitudes of $n_{\parallel,\perp}$, or of their
average, must be done rather than of the difference $n_{\parallel} - n_{\perp}$, so that birefringence experiments
such as PVLAS cannot be used (although birefringence can and has been used to place bounds on more general nonlinear
electromagnetic theories \cite{pvlasphotonphoton}).  
As discussed by various authors \cite{debvan,dekrkr,hlassw,zavcal,doegie}, the measurement of the QED refraction indices may soon become viable 
using large scale laser interferometers such as LIGO, GEO or VIRGO, simply by inducing interference through the
application of a strong magnetic field perpendicularly to one of the two legs of the interferometer. 
From (\ref{nQED}) and (\ref{ntotal}) one finds that the confirmation of the QED
refraction indices would lead to a lower bound on $b$ roughly equal to the value
of the original Born-Infeld theory, (\ref{defb}) (this fact is still related
to Infeld's observation about the QED and BI four-point vertices mentioned in the introduction).

Coming to the case of photon splitting, a definite observation of this process so far was achieved only for the electric field case, 
and again using the Coulomb field of heavy atoms \cite{akhmadalievetal}. 
However, this now corresponds to a calculation entirely different from the one above since the
box diagram does not drop out in the Coulomb field case, so that the inclusion of
derivative corrections becomes even more imperative than for Delbr\"uck scattering.
Keeping thus to the (constant) magnetic field case, due to the factor of $(B/B_{\rm cr})^6$ in (\ref{kappatotal})
here measurable effects can be expected only for magnetic fields close to $B_{\rm cr}$; 
in laboratory experiments such as  PVLAS magnetic photon splitting in principle contributes to dichroism, but the effect is tiny.
Thus the natural application of the magnetic effect, which was in fact already the motivation for Adler's original
calculation in 1971 \cite{adler71}, is to the physics of neutron stars, which are known to have surface magnetic fields 
close to or even exceeding $B_{\rm cr}$ (for a recent review see \cite{mereghetti}). 
Photon splitting could be seen in the spectra of neutron stars both
through its softening and its polarizing effect, and, although no definite detection seems to have been reported as of date,
recent results from magnetars are suggestive of such a softening effect \cite{enoto_ea}.  
If it could be ascertained that this effect is due to photon splitting, and assuming that, as a result, the QED contribution to
the absorption coefficient (\ref{kappatotal}) would be confirmed with a (say) 10 percent error margin, then the lower
bound on $b$ from (\ref{kappatotal}) would come out as

\bear
b &\geq& 2.0 \times 10^{19} \frac{V}{m} \, .
\label{bsplitting}
\ear
Although the study of \cite{enoto_ea} concerns hard x-rays, due to the absence of derivative corrections to the hexagon diagram 
here this does not lead us out of the range of the applicability of the low photon-energy approximation. 
For $B/B_{\rm cr} \sim 1$ higher-point corrections can be sizable, but according to the numerical studies of \cite{adler71,bamish} 
at least in the QED case do not tend to reduce the  hexagon contribution (this is in contrast to the Coulomb
field photon splitting where it has been shown that the leading four-point contribution significantly overestimates
the amplitude \cite{lemist}). 

Thus the bound (\ref{bsplitting}) would again be close to the Born-Infeld value (\ref{defb}), and still be three orders of magnitude 
below the one achieved by Soff et al. \cite{soragr}. However, it must be kept
in mind that the latter was derived in fundamental Born-Infeld theory, without
contemplating possible derivative corrections. The example of Delbr\"uck 
scattering in a Coulomb field discussed above leads us to expect that the inclusion
of such corrections in the set-up of \cite{soragr} may lead to substantive changes.
The bound from magnetic photon splitting would have the advantage of being 
relatively insensitive to possible derivative corrections, both with respect to the
field photons (since the magnetic field involved changes only on macroscopic scales)
and the splitting ones (due to the fortuitous kinematics of the process which makes
the leading hexagon approximation coincide with its low-energy limit).

\section{Conclusions}
\label{conclusions}
\renewcommand{\theequation}{5.\arabic{equation}}
\setcounter{equation}{0}

To summarize, we have obtained here the following tree-level quantities in Born-Infeld theory:
the total photon-photon scattering cross section, the photon splitting amplitude in a constant
magnetic field in the leading (hexagon) approximation, and also the related refractive indices
in a weak constant field. We have also shown that Adler's selection rules for magnetic photon
splitting in QED hold for the Born-Infeld case unchanged.

Each of these tree-level quantities has a one-loop analogue in QED,
and can in principle be used for constraining the Born-Infeld parameter $b$. Discussing  various
experimental options we have come to the conclusion that what is achievable along these
lines in the near future is a lower bound on $b$ close to the value of the original Born theory,
eq. (\ref{defb}). Presently photon splitting appears to be the most promising one of these processes.

Finally, we would like to point out that photon splitting may turn out to be
particularly useful for testing more general non-linear electromagnetic theories, since for those contrary to the Born-Infeld case
Adler's selection rules may be violated, thus opening up photon splitting channels which are forbidden in QED.

\medskip
\noindent
{\bf Acknowledgements:}
We are obliged to D. Bernard, I. Bialynicka-Birula, H. Gies, T. Heinzl, K. Makishima, M. Reuter, C. Sonnenschein, J.B. Zuber
and B. Zwiebach  for various helpful discussions or correspondence. Special thanks to D. Bernard
and H. Gies for useful comments on the manuscript. 
All three authors thank CONACYT for financial support.

%
%


\end{document}